\documentclass[letterpaper]{aipproc}
\usepackage[hypertex]{hyperref}

\makeatletter
\newcommand{\vereq}[2]{%
 \lower3\p@\vbox{%
  \baselineskip1.5\p@
  \lineskip1.5\p@
  \ialign{$\m@th#1\hfill##\hfil$\crcr#2\crcr\sim\crcr}%
 }%
}%
\newcommand{\lesssim}{\mathrel{\mathpalette\vereq{<}}}%
\makeatother

\layoutstyle{6x9}

\newcommand\doingARLO[2][]{%
  \ifx\mmref\undefined #1\else #2\fi
}

\begin{document}

\title[Impact on Theory]{Impact of Neutrino Oscillation\\ Measurements
  on Theory} 


\author{Hitoshi Murayama}{
  address={School of Natural Sciences, Institute for Advanced Study,
  Princeton, NJ 08540},
  email={murayama@ias.edu},
  homepage={http://hitoshi.berkeley.edu},
  altaddress={Department of Physics, University of California,
    Berkeley, CA 94720},
  thanks={This work was supported by the Institute for Advanced Study,
    funds for Natural Sciences, as well as in part by the DOE under
    contracts DE-FG02-90ER40542 and DE-AC03-76SF00098 and in part by
    NSF grant PHY-0098840.}  }

\copyrightyear  {2003}

\begin{abstract}
  Neutrino oscillation data had been a big surprise to theorists, and
  indeed they have ongoing impact on theory.  I review what the impact
  has been, and what measurements will have critical impact on theory
  in the future.
\end{abstract}

\date{\today}

\maketitle

\section{Introduction}

I was asked to comment on the impact of neutrino oscillation
measurements on theory.  It is completely clear that recent neutrino
oscillation data had big impact on theory, and it will continue to do
so.  I will remind you about the ongoing impact.  Then I will list
measurements that will have critical impact on theory in the future.

Let me organize my discussion as ``Past,'' ``Present,'' and
``Future.''

\section{Past}

It is useful to recall why theorists had always been interested in the
small neutrino masses and their consequences on neutrino oscillation.
It is because we are always interested in probing physics at as high
energies as possible.  One way to probe it is of course to go to the
high-energy collider experiments and study physics at the energy scale
directly.  Another way is to look for rare and/or tiny effects coming
from the high-energy physics.  The neutrino mass belongs to the second
category.

To study rare and/or tiny effects from physics at high energies, we
can always parameterize them in terms of the power series expansion,
\begin{equation}
  {\cal L} = {\cal L}_4 + \frac{1}{\Lambda} {\cal L}_5 +
  \frac{1}{\Lambda^2} {\cal L}_6 + \cdots.
\end{equation}
The zeroth order term ${\cal L}_4$ is renormalizable and describes the
Standard Model.  On the other hand, the higher order terms are
suppressed by the energy scale of new physics $\Lambda$.  Possible
operators can be classified systematically, which I believe was done
first by Weinberg (but I couldn't find the appropriate reference).
With two powers of suppression, there are many terms one can study:
\begin{equation}
  {\cal L}_6 \supset QQQL,\, \bar{L} \sigma^{\mu\nu} W_{\mu\nu} H e,\,
  W_\nu^\mu W_\lambda^\nu B_\mu^\lambda,\,
  \bar{s}d \bar{s}d,\, (H^\dagger D_\mu H) (H^\dagger D^\mu H), \cdots
\end{equation}
The examples here contribute to proton decay, $g-2$, anomalous triple
gauge boson vertex, $K^0$--$\overline{K}^0$ mixing, and the
$\rho$-parameter, respectively.  It is interesting that there is only
one operator suppressed by a single power:
\begin{equation}
  {\cal L}_5 = (LH) (LH).
\end{equation}
After substituting the expectation value of the Higgs, the Lagrangian
becomes
\begin{equation}
  {\cal L} = \frac{1}{\Lambda} (LH)(LH)
  \rightarrow \frac{1}{\Lambda} (L\langle H\rangle)(L\langle H\rangle)
  = m_\nu \nu \nu,
\end{equation}
nothing but the neutrino mass.

Therefore the neutrino mass plays a very unique role.  It is the
lowest-order effect of physics at short distances.  This is a very
tiny effect.  Any kinematical effects of the neutrino mass are
suppressed by $(m_\nu / E_\nu)^2$, and for $m_\nu \sim 1$~eV which we
now know is already too large and $E_\nu \sim 1$~GeV for typical
accelerator-based neutrino experiments, it is as small as $(m_\nu /
E_\nu)^2 \sim 10^{-18}$.  At the first sight, there is no hope to
probe such a small number.  However, any physicist knows that
interferometry is a sensitive method to probe extremely tiny effects.
For interferometry to work, we need a coherent source.  Fortunately
there are many coherent sources of neutrinos in Nature, the Sun,
cosmic rays, reactors (not quite Nature), etc.  We also need
interference for an interferometer to work.  Because we can't build
half-mirrors for neutrinos, this could have been a show stopper.
Fortunately, there are large mixing angles that make the interference
possible.  We also need long baselines to enhance the tiny effects.
Again fortunately there are many long baselines available, such as the
size of the Sun, the size of the Earth, etc.  Nature was very kind to
provide all necessary conditions for interferometry to us!  Neutrino
interferometry, a.k.a.  neutrino oscillation, is therefore a unique
tool to study physics at very high energy scales.

\begin{figure}[t]
  \includegraphics[width=0.7\textwidth]{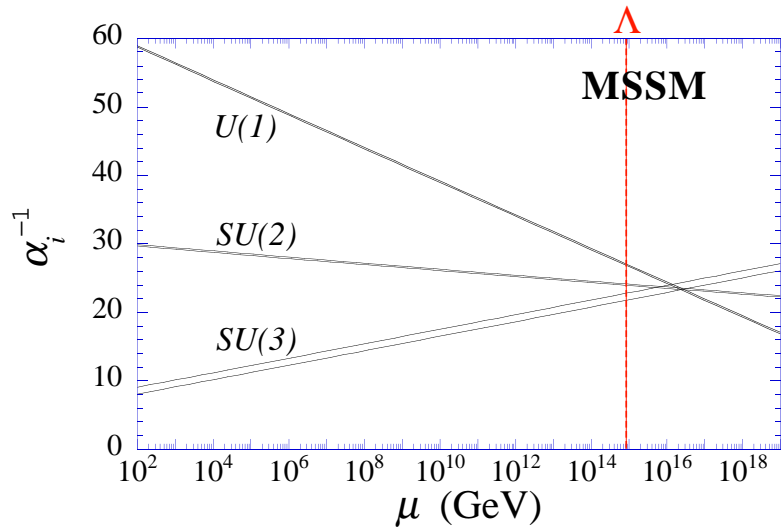}
  \caption{Apparent unification of gauge coupling unification in the
    MSSM at $2 \times 10^{16}$~GeV, compared to the suggested scale of
    new physics from the neutrino oscillation data.\label{fig:MSSM}}
\end{figure}

Indeed, the recently established neutrino oscillation results
\cite{Shiozawa,Nakamura}
\begin{eqnarray}
  \Delta m_{\it atm}^2 &\sim& 0.002 {\rm eV}^2, \\
  \Delta m_{\it solar}^2 &\sim& 0.00007 {\rm eV}^2,
\end{eqnarray}
interpreted naively in a ``hierarchical'' mass scheme
\begin{eqnarray}
  m_3 &\sim& \sqrt{\Delta m_{\it atm}^2} \sim 0.04 {\rm eV}, \\
  m_2 &\sim& \sqrt{\Delta m_{\it solar}^2} \sim 0.008 {\rm eV},
\end{eqnarray}
suggests
\begin{equation}
  \Lambda \sim \frac{\langle H \rangle^2}{m_3} \sim 8 \times
  10^{14}~{\rm GeV}.
\end{equation}
It is tantalizingly close to the energy scale of apparent gauge
coupling unification in the Minimal Supersymmetric Standard Model, $2
\times 10^{16}$~GeV.  (See, Fig.~\ref{fig:MSSM}.)

This way, the neutrino oscillation appears to provide us a unique
window to physics at very high energies as their ``leading order''
effects.  Indeed, theoretical estimates based on the seesaw mechanism
in the grand unified theories \cite{seesaw} are practically confirmed!

\section{Present}

\begin{figure}[t]
  \centering{
    \includegraphics[width=0.5\textwidth]{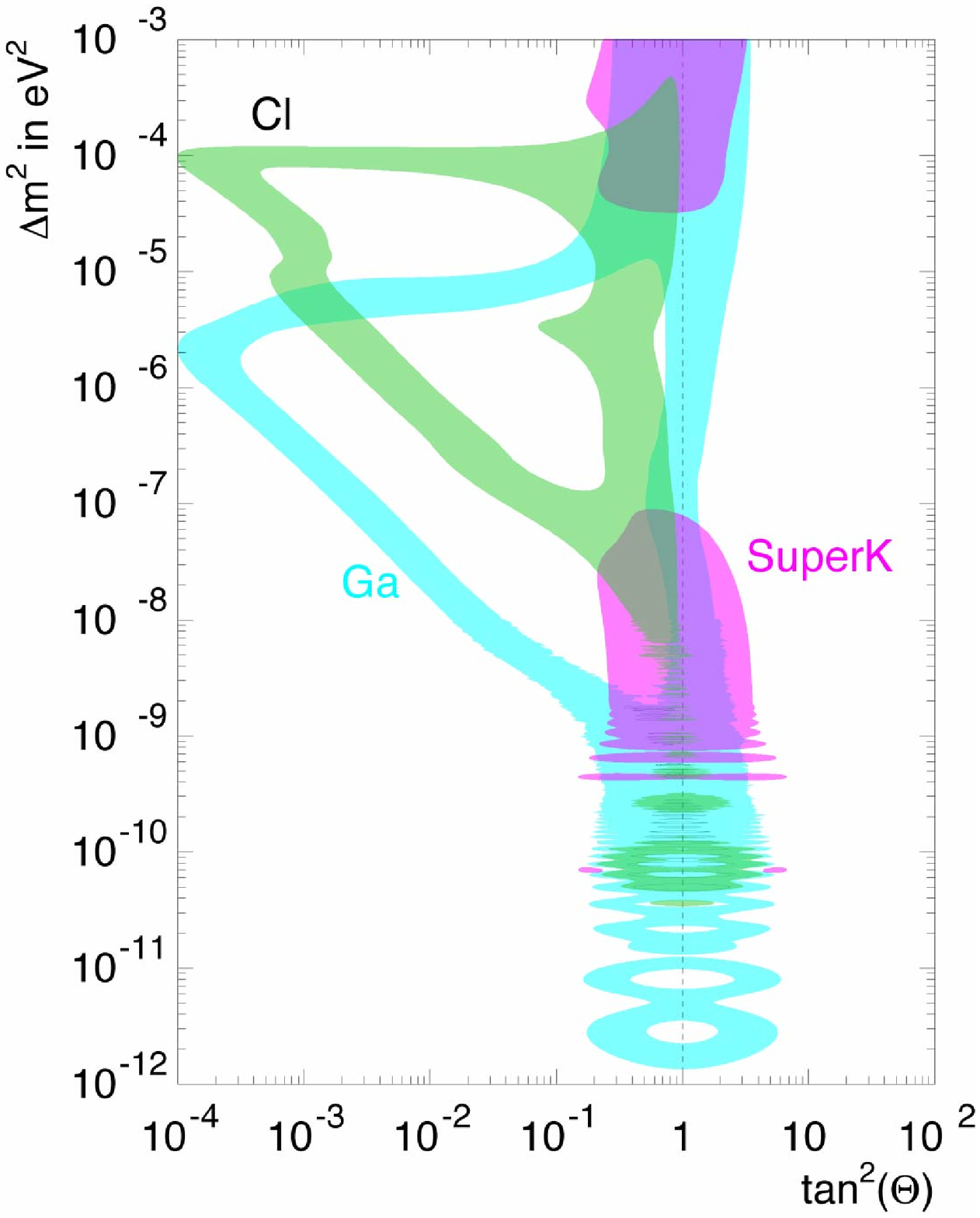}
    \includegraphics[width=0.5\textwidth]{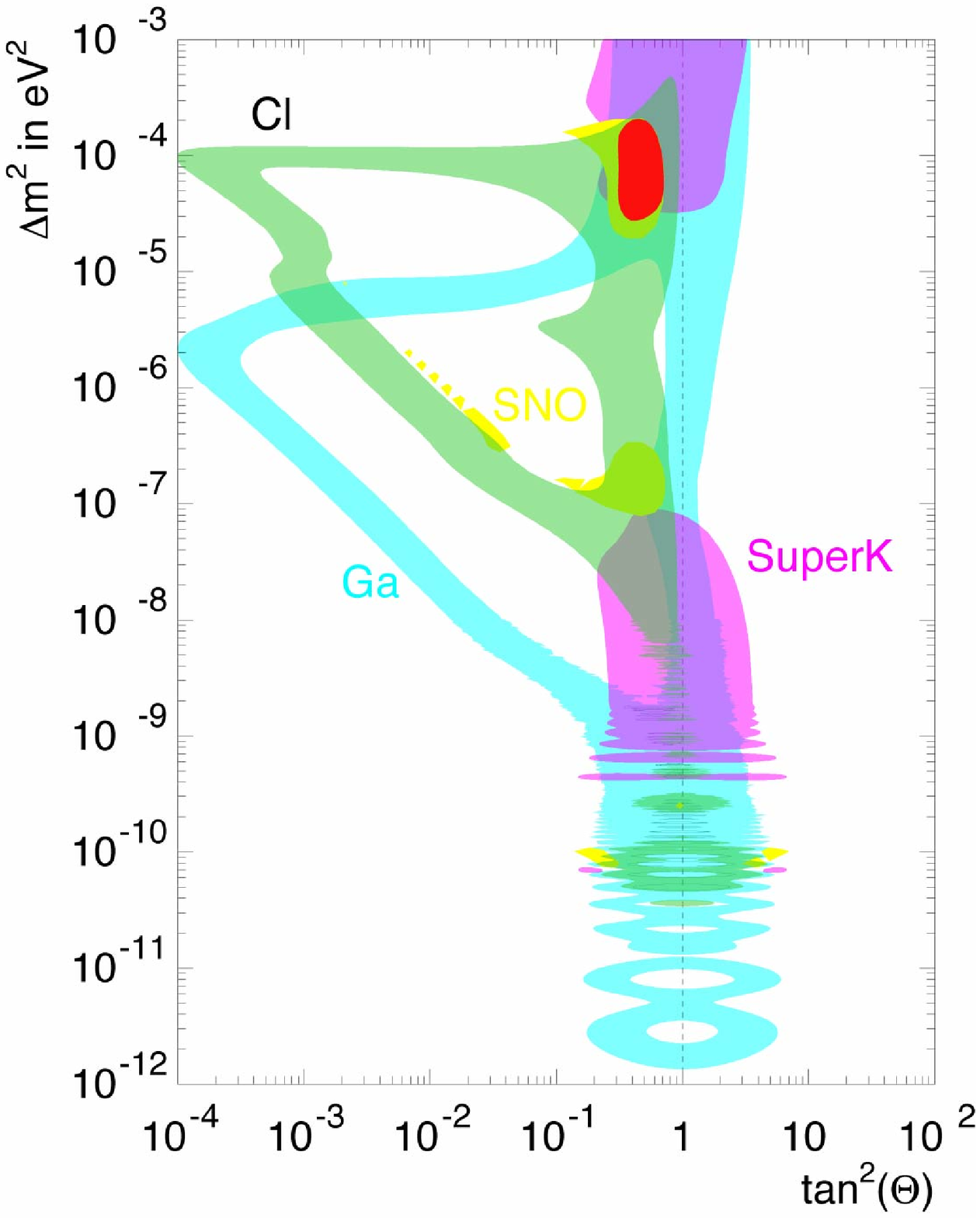}}
  \caption{The progress in solar neutrino in the year 2002.  Before
    March and after April \cite{hitoshi}.\label{fig:allsolar1} }
\end{figure}

\begin{figure}[t]
\centering{
    \includegraphics[width=0.5\textwidth]{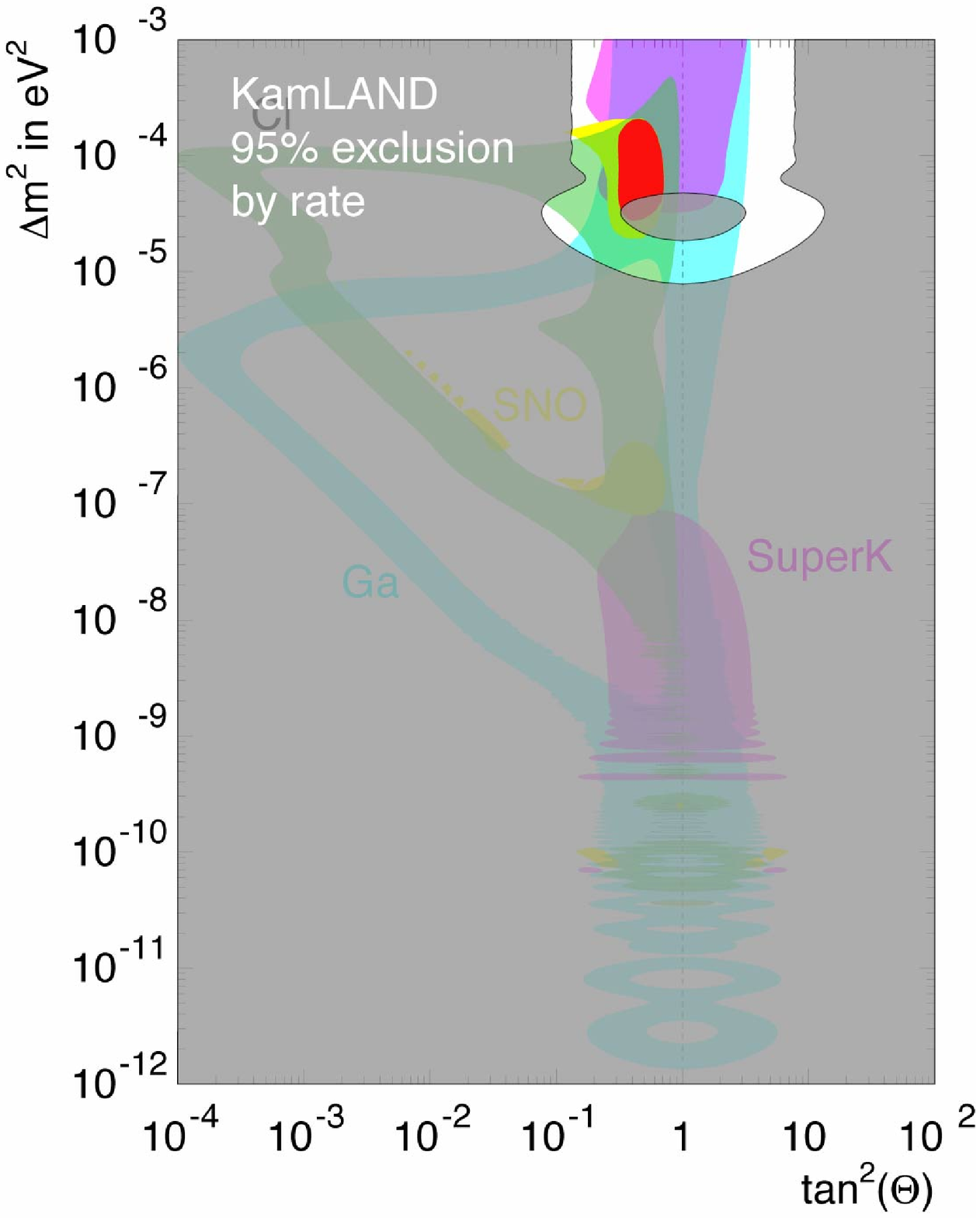}
    \includegraphics[width=0.5\textwidth]{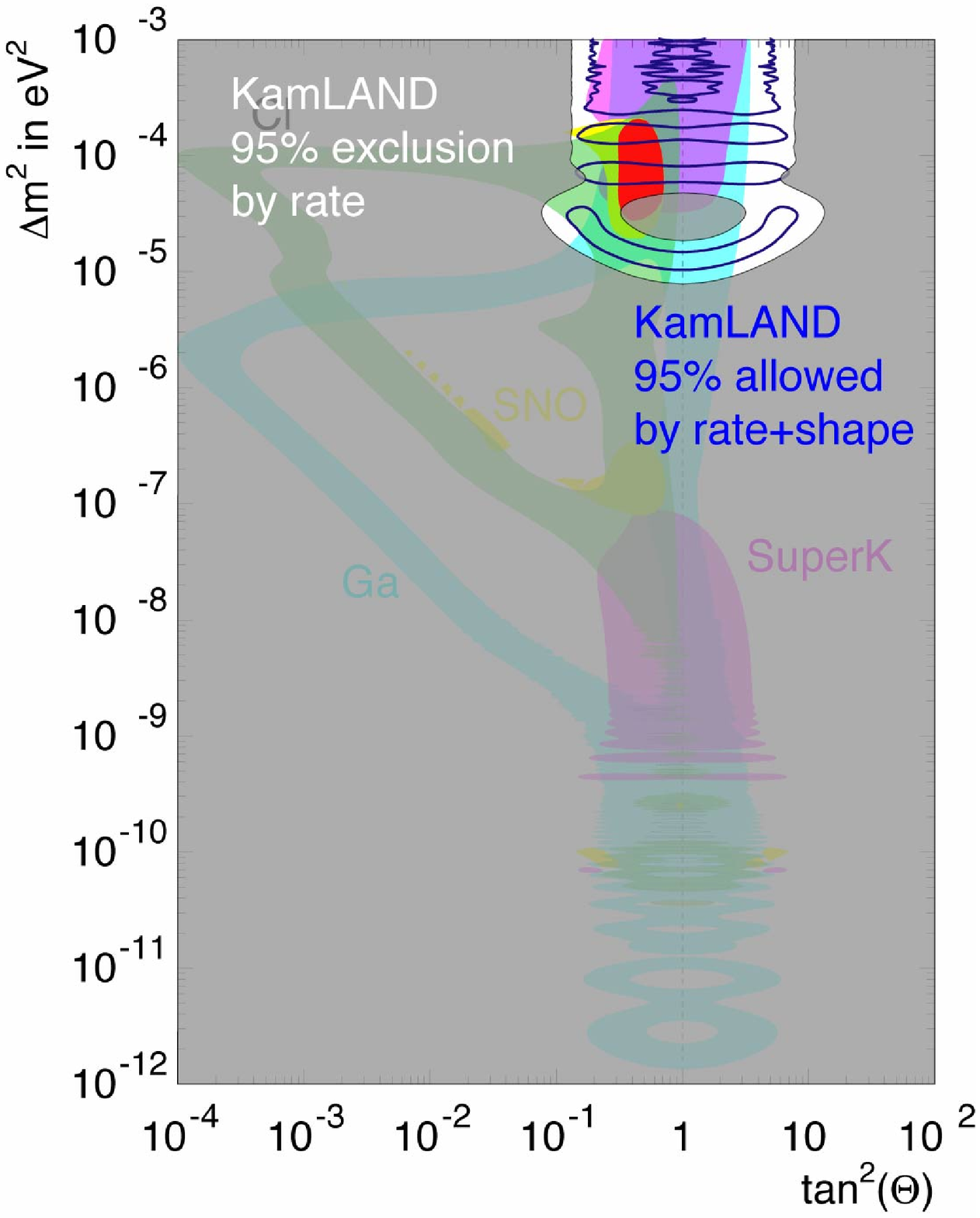}}
  \caption{The comparison of the solar neutrino data and the reactor
    anti-neutrino data after December
    \cite{hitoshi}.\label{fig:allsolar2}}
\end{figure}

The last year was an amazing year in neutrino physics.  Before March,
the situation of the solar neutrino data looked like the first plot in
Fig.~\ref{fig:allsolar1}, and there had been overlaps between SuperK,
Homestake, and Gallium experiments in the LMA and LOW regions, some
down in quasi-vaccum.  After SNO neutral current result in April, the
parameter space focused only on the LMA region shown in red in the
second plot in Fig.~\ref{fig:allsolar1}.  In December, KamLAND has
excluded most of the parameter space as shown in the first plot in
Fig.~\ref{fig:allsolar2}, while its preferred region (inside the blue
contours in the second plot in Fig.~\ref{fig:allsolar2}) has
consistent overlaps with the that preferred by the solar neutrino
data.  It was a tremendous convergence from the parameter space over
many decades down to factors of a few.

It is useful to recall what a typical theorist used to say back around
1990.  
\begin{itemize}
\item The solution to the solar neutrino problem must be the small
  mixing angle MSW solution because it is so beautiful.
\item The natural scale for $\nu_\mu \rightarrow \nu_\tau$ oscillation
  is $\Delta m^2 \sim {\rm eV}^2$ because it is the cosmologically
  interesting range.
\item The angle $\theta_{23}$ must be of the same order of magnitude
  as $V_{cb}$ because of the grand unification.
\item The atmospheric neutrino anomaly must go away because it would
  require a large mixing angle to explain.
\end{itemize}
Needless to say, theorists have a very good track record in neutrino
physics.

Indeed, the recent results from neutrino oscillation physics had
surprised almost everybody.  The prejudice has been that the mixing
angles must be small because quark mixing angles are small, and the
masses must be hierarchical because both quarks and charged lepton
masses are hierarchical.  Given that the LMA is now chosen, all mixing
angles are large except for $U_{e3}$ that must be small-ish (but the
current limit is not very strong, $|U_{e3}| \lesssim 0.2$).

The natural question then is if this newly discovered surprising
pattern of neutrino masses and mixings require a new symmetry or any
special structure to explain.  

In fact, the big question has always been {\it what distinguishes
  flavor?}\/ Three generations share exactly the same quantum numbers.
Yet, they have such different masses.  The hierarchy with small
mixings means that there is a need for some kind of ordered structure.
The ``common sense'' in quantum mechanics is that states with the same
quantum numbers should have similar energy levels ({\it i.e.}\/
masses) and mix significantly under small perturbations.  The observed
patterns go against this ``common sense.''  The hierarchical masses
and small mixings among quarks and charged leptons had been a puzzle.

Therefore, there has been a strong suspicion that there is a new set
of quantum numbers, {\it flavor quantum numbers}\/, that distinguish
three generations of quarks and leptons.  As Noether told us, a new
quantum number requires a new symmetry, {\it flavor symmetry}\/.  This
new symmetry must allow the top quark Yukawa coupling because it is of
the natural size, $y_t \simeq 1.0$.  On the other hand, all the other
Yukawa couplings are practically zero (as opposed to $O(1)$), and the
flavor symmetry must forbid them.  After the symmetry is broken by a
small parameter, all the other Yukawa couplings become allowed, but
suppressed \cite{Froggatt:1978nt}.  The hope is to identify the
underlying symmetry based on the data, similarly to what was done by
Heisenberg (isospin) or Gell-Mann--Okubo (flavor $SU(3)$).

Indeed, the neutrino data had been already effective in narrowing down
the possibilities of flavor symmetries.  In Table \ref{tab:AFM}, many
proposed flavor symmetries are shown together with their predictions
on the mass-squared ratio $\Delta m^2_{12}/\Delta m^2_{23}$, $U_{e3}$,
$\theta_{12}$, and $\theta_{23}$, taken from \cite{Altarelli:2002sg}
(October 2002).\footnote{$d_{23}$ in the model {\tt SA} refers to a
  degree of accidental cancellation in the 23 sector, that is used to
  enhance $\theta_{12}$.}  Since then, models {\tt H$_{\tt II}$}, {\tt
  H$_{\tt I}$}, and {\tt IH (LOW)} had been excluded by KamLAND.

\begin{table}[t]
\caption{Prediction of different flavor symmetries on the neutrino
  mass-squared ratio and various mixing angles, taken from
  \cite{Altarelli:2002sg}.\label{tab:AFM}} 
\centering
\begin{tabular}{|c|c|c|c|c|c|c|}
\hline
Model& parameters& $d_{23}$ & $\Delta m^2_{12}/\vert\Delta
m^2_{23}\vert$& $U_{e3}$& $\tan^2\theta_{12}$ &$\tan^2\theta_{23}$\\ 
\hline
{\tt A} &$\epsilon=1$ & O(1) & O(1) & O(1) & O(1) & O(1) \\
\hline
{\tt SA} &$\epsilon=\lambda$ & O(1) & O($d_{23}^2$) & O($\lambda$) &
O($\lambda^2/d_{23}^2$)  & O(1) \\ 
\hline
{\tt H$_{\tt II}$} &$\epsilon=\lambda^2$ & O($\lambda^2$) &
O($\lambda^4$) & O($\lambda^2$) &  O(1)  & O(1) \\
\hline
{\tt H$_{\tt I}$} &$\epsilon=\lambda^2$ & 0 & O($\lambda^6$) &
O($\lambda^2$) &  O(1)  & O(1) \\
\hline
{\tt IH (LA)} &$\epsilon=\eta=\lambda$ & O($\lambda^4$) &
O($\lambda^2$) & O($\lambda^2$) &  1+O($\lambda^2$)  & O(1) \\
\hline
{\tt IH (LOW)} &$\epsilon=\eta=\lambda^2$ & O($\lambda^8$) &
O($\lambda^4$) & O($\lambda^4$) &  1+O($\lambda^4$)  & O(1) \\
\hline
\end{tabular}
\end{table}

Among them, I liked the model {\tt A} the best, because it is {\it
  mine}\/ \cite{Hall:1999sn}.  It is called {\it anarchy}\/, based on
the idea that neutrinos are actually normal, while quarks and charged
leptons aren't.  As I mentioned already, the hierarchical masses and
small mixing angles are against the ``common sense,'' while the
neutrinos do not seem to have a large hierarchy and mix a lot.  Maybe
the {\it lack of flavor symmetry}\/ can explain the data.  Indeed, if
there is no fundamental distinction among three neutrinos, or in other
words if their flavor quantum numbers are all equal, the group theory
of three-by-three unitary matrices uniquely determine the probability
distribution of mixing angles \cite{Haba:2000be}.  Then all three
angles, $\theta_{12}$, $\theta_{23}$, and $\theta_{13}$ are three
random draws from the distribution $dP/dx \propto (1-x)^{-1/2}$ for
$x=\sin^2 2\theta$.  Because it is peaked towards the maximal angle
$x=1$, it is very plausible that two draws come out large, while one
of them comes down the tail (but not expected {\it way}\/ down the
tail).  Indeed, the Kolgomorv--Smirnov test suggests that the
probability that three random draws come out worse than the actual
data is 64\%, and hence the observed pattern is completely natural if
there is {\it no}\/ fundamental distinction among three generations
\cite{deGouvea:2003xe}.  On the other hand, $\theta_{13}$ is expected
to be not too far below the current limit.  The one-dimensional KS
probability is $P(KS) = 4 (\sin^2 \theta_{13} - \frac{1}{2} \sin^4
\theta_{13})$, and hence we expect $\sin^2 \theta_{13} > 0.013$ at
``95\% CL.''  The size of the CP-violation $\sin \delta$ is
distributed as $1/|\cos\theta|$, and hence is expected to be large.

\begin{figure}[t]
  \centering
  \includegraphics[width=0.5\textwidth]{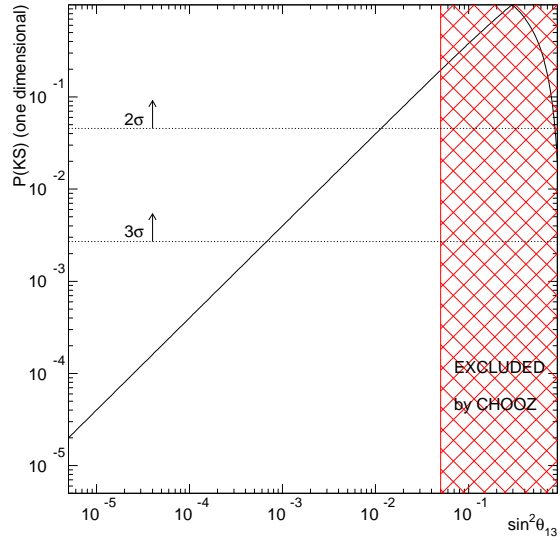}
  \caption{The one-dimensional KS probability on $\sin^2\theta_{13}$ based on
    anarchy, {\it i.e.}\/, no fundamental distinction among three
    neutrinos.  Taken from \cite{deGouvea:2003xe}.}
  \label{fig:KS}
\end{figure}

\section{Future}

Having discussed the impact of neutrino data on theory so far, it is
clear what will be the critical measurements in the future.  
\begin{itemize}
\item $\sin^2 2\theta_{23} = 1.00 \pm 0.01$?  If it comes out that
  precisely maximal, it surely will require a new symmetry.
\item $\sin^2 \theta_{13} < 0.01$?  If so, electron-neutrino must have
  a different flavor quantum number from muon and tau neutrinos.
\item Normal or inverted hierarchy?  Most flavor symmetries predict
  the normal hierarchy, but theorists had been wrong!
\item CP Violation?  Even though the CP violation in neutrino
  oscillation may not prove the relevant CP violation for
  leptogenesis, it will at least make it very plausible.
\end{itemize}

After going through the critical measurements, we hope to determine
the underlying flavor symmetries behind neutrinos (and flavor in
general).  Then comes an even bigger question: can we understand the
dynamics behind the flavor symmetry?  In the case of the strong
interaction, isospin and flavor $SU(3)$ are the flavor symmetries,
while the QCD is the dynamics.  Can we get to the same level?  This
question will depend crucially on what we will find at the TeV-scale.
If it is supersymmetry, the answer may be anomalous U(1) gauge
symmetry with the Green--Schwarz mechanism from the string theory
\cite{anomalousU(1)}.  If it is extra dimensions, the answer may be
physical dislocation of different particles within a thick brane
\cite{Arkani-Hamed:1999dc}.  If it is technicolor, the answer may be
new broken gauge symmetries at 100~TeV scale \cite{Eichten:1979ah}.

Of course one shouldn't forget LSND \cite{Louis} because no theory
fit the data very well.  It is true that most theorists do not take
the LSND evidence seriously at this moment, only data will decide.
Currently all explanations have difficulties: sterile neutrino(s)
\cite{Maltoni:2002xd}, CPT violation \cite{Gonzalez-Garcia:2003jq},
lepton-number violating muon decay \cite{Armbruster:2003pq}.  But if
any of them will turn out to be true, it will have a huge impact on
theory.

\section{Conclusion}

Neutrino oscillation physics has had big impact on theory already.
Yet, there is a lot more to learn.  The (precise) measurements of
$\theta_{23}$, $\theta_{13}$, the type of hierarchy, and the CP
violation, will have critical impact.  Through these measurements, we
hope to determine the symmetries behind the neutrino masses and
mixings or flavor in general.  In conjunction with data from the
energy frontier, we may even have access to understand dynamics behind
the flavor.  Depending on how things will turn out, there may well be
even more surprises.  

\begin{theacknowledgments}
  This work was supported by the Institute for Advanced Study, funds
  for Natural Sciences, as well as in part by the DOE under contracts
  DE-FG02-90ER40542 and DE-AC03-76SF00098 and in part by NSF grant
  PHY-0098840.
\end{theacknowledgments}



\end{document}